\documentclass[usenatbib]{aa}  
\usepackage{natbib,twoopt}
\usepackage{graphicx}
\usepackage{txfonts}
\usepackage{academicons}
\usepackage{hyperref}
\usepackage{cleveref}
\usepackage[T1]{fontenc}
\usepackage[dvipsnames]{xcolor}
\usepackage[normalem]{ulem}

\begin{document}

   \title{The galaxy bias profile of cosmic voids}

   \titlerunning{The bias profile inside cosmic voids}

    \authorrunning{Montero-Dorta et al.}

   \author{Antonio D. Montero-Dorta\inst{1}\thanks{antonio.montero@usm.cl}, Andrés Balaguera-Antolínez$^{2,3}$, Ignacio G. Alfaro$^{4,5}$, Andrés N. Ruiz$^{4,5}$, Ravi K. Sheth$^{6,7}$, Facundo Rodriguez$^{4,5}$, Daniela Galárraga-Espinosa$^{8,9}$, Constanza A. Soto-Suárez$^{10,11}$, Ignacio Quiroz$^{10,11}$, Iker Fernández-Sánchez$^{2,3}$
           }

   \institute{
             Departamento de F\'isica, Universidad T\'ecnica Federico Santa Mar\'ia, Avenida Vicu\~na Mackenna 3939, San Joaqu\'in, Santiago, Chile \and 
 Instituto de Astrof\'{\i}sica de Canarias, s/n, E-38205, La Laguna, Tenerife, Spain \and 
 Departamento de Astrof\'{\i}sica, Universidad de La Laguna, E-38206, La Laguna, Tenerife, Spain \and CONICET. Instituto de Astronomía Teórica y Experimental (IATE). Laprida 854, Córdoba X5000BGR, Argentina
   \and
        Universidad Nacional de Córdoba (UNC). Observatorio Astronómico de Córdoba (OAC). Laprida 854, Córdoba X5000BGR, Argentina \and Center for Particle Cosmology, University of Pennsylvania, Philadelphia, PA 19104, USA \and The Abdus Salam International Center for Theoretical Physics, Strada Costiera 11, Trieste 34151, Italy \and Kavli IPMU (WPI), UTIAS, The University of Tokyo, Kashiwa, Chiba 277-8583, Japan
        \and
        Max Planck Institute for Astrophysics, Karl-Schwarzschild-Straße 1, D-85748 Garching, Germany
        \and Departamento de F\'isica, Universidad T\'ecnica Federico Santa Mar\'ia, Casilla 110-V, Avda. Espa\~na 1680, Valpara\'iso, Chile \and Instituto de Física, Pontificia Universidad Católica de Valparaíso, Casilla 4950, Valparaíso, Chile
             }

   \date{Received --; accepted --}

 
  \abstract
   {Cosmic voids are underdense regions within the large-scale structure of the Universe, spanning a wide range of physical scales -- from a few megaparsecs (Mpc) to the largest observable structures. Their distinctive properties make them valuable cosmological probes and unique laboratories for galaxy formation studies. A key aspect to investigate in this context is the galaxy bias, $b$, within voids -- that is, how galaxies in these underdense regions trace the underlying dark-matter density field.}
   {We want to measure the dependence of the large-scale galaxy bias on the distance to the void center, and to evaluate whether this bias profile varies with the void properties and identification procedure.
   }
   {We apply a void identification scheme based on spherical overdensities to galaxy data from the IllustrisTNG magnetohydrodynamical simulation. For the clustering measurement, we use an object-by-object estimate of large-scale galaxy bias, which offers significant advantages over the standard method based on ratios of correlation functions or power spectra.}
   {We find that the average large-scale bias of galaxies inside voids tends to increase with void-centric distance when normalized by the void radius. For the entire galaxy population within voids, the average bias rises with the density of the surrounding environment and, consequently, decreases with increasing void size. Due to this environmental dependence, the average galaxy bias inside S-type voids -- embedded in large-scale overdense regions -- is significantly higher ($\langle b\rangle_{\rm in} > 0$) at all distances compared to R-type voids, which are surrounded by underdense regions ($\langle b\rangle_{\rm in} < 0$). The bias profile for S-type voids is also slightly steeper. Since both types of voids host halo populations of similar mass, the measured difference in bias can be interpreted as a secondary bias effect.}
   {}


   \maketitle
%

\section{Introduction}

The large-scale structure of the Universe (LSS) can be described as a complex web of filaments and plane-like dark matter structures that intersect to form massive virialized systems -- regions where galaxy formation is most likely to occur. This understanding is based on the prevailing model of structure formation, which posits that galaxies form as baryonic matter cools and condenses within the gravitational potential wells shaped by the collisionless collapse of dark matter halos \citep{White1978}. However, the diverse astrophysical processes governing galaxy formation and evolution introduce significant uncertainty in both the properties of galaxies and their distribution within halos. Deciphering the connection between galaxies and their dark matter halos is essential for understanding the formation and evolution of large-scale cosmic structures \citep{Wechsler2018}.

As matter accumulates in some regions of the Universe, large-scale underdense regions -- known as cosmic voids -- naturally emerge elsewhere. Voids represent the lowest-density regions of the cosmic web, where overdensities are minimal relative to the mean background density of the Universe. The shape and internal structure of voids have been shown to be complex (e.g., \citealt{Gottloeber2003, AragonCalvo2013}), as demonstrated by the variety of identification schemes proposed in the literature (e.g., \citealt{Hahn2007,Platen2007, Neyrinck2008,Lavaux2010,Sutter2015,Elyiv2015,Ruiz2015,Paz2023}), often yielding differing results (e.g., \citealt{Colberg2008, Cautun2018}). 

Cosmic voids occupy the majority of the Universe's volume but contain only a small fraction of the galaxy population \citep{Pan2012}. One of the simplest ways to define them is as expanding spherical regions with densities of 10 to 20\% the mean density of the Universe \citep{Sheth_voids, Ceccarelli2006, Ruiz2015}. These extremely low-density conditions result in a distinct halo population inside voids compared to that inhabiting higher-density regions \citep{Mo1996,ShethTormen2002}. As a result, galaxies residing in voids tend to be fainter, bluer, and predominantly of late-type morphologies \citep{Rojas2004, Hoyle2005, Patiri2006, Ceccarelli2008, Hoyle2012}. They typically host younger stellar populations and exhibit higher star formation rates \citep{Rojas2005}. The low mass density and faster expansion of voids produce also a local Hubble constant larger than the cosmic average \citep{Weygaert1993, Tomita2000}, which, along with the absence of rich clusters, tend to inhibit structure growth. This physical context also shapes the halo-galaxy connection, as halos in voids, particularly the most massive ones, host significantly fewer satellite galaxies than those in denser regions \citep{Alfaro2020, Alfaro2022}. These results highlight fundamental differences in the way galaxies populate halos across cosmic environments.

As the largest and least dense structures in the Universe, cosmic voids serve as valuable cosmological probes, currently used to constrain cosmological models and structure formation theories. Key statistics measured in this context include void number counts, clustering, and the void size function (e.g., \citealt{Hamaus2015, Pisani2015, Achitouv2017, Chuang2017, Hamaus2017, Hawken2017, Achitouv2019, Kreisch2019, Nadathur2019, Verza2019, Hamaus2020, Nadathur2020, Bayer2021, Correa2021, Kreisch2022, Contarini2022, Moresco2022, Woodfinden2022, Contarini2023,Song2024}). These and other void-related measurements can be used to study dark energy, modified gravity (e.g., \citealt{Biswas2010, Lavaux2010, Clampitt2013, Pisani2015, Zivick2015, Pollina2016, Sahlen2016, Falck2018, Paillas2019, Perico2019, Verza2019, Contarini2021, Contarini2022, Verza2023}), and primordial non-Gaussianity \citep{Chan2019}, among other open questions in the field. The role of cosmic voids in cosmology is expected to grow with current and upcoming surveys, such as the Dark Energy Spectroscopic Instrument (DESI; \citealt{DESI2016}), Prime Focus Spectrograph (PFS; \citealt{Tamura2016}), Euclid \citep{Laureijs2011}, and the Vera C. Rubin Observatory \citep{Ivezic2019}, among others.

Understanding how galaxies and halos trace the underlying dark-matter density field is a fundamental goal in cosmology. This relationship, often characterized in terms of the so-called \emph{bias} ($b$), has been extensively studied in the context of voids (\citealt{Furlanetto2006, Hamaus2014, Sutter2014A, Sutter2014B, Neyrinck2014, Clampitt2016, Pollina2016, Pollina2017, Pollina2019, Contarini2019, Voivodic2020, Verza2022, Verza2023, Verza2024}). It is commonly assumed that a simple linear relation between tracer and matter overdensity exists across scales inside voids (i.e., for halos, $\delta_{\rm h}(r) \simeq b_{\rm h} \delta_{\rm m}(r)$, with the possible exception of a small additive constant; see \citealt{Pollina2017, Pollina2019,Verza2022}). However, recent findings challenge this assumption. In particular, \cite{Verza2022} used cosmological simulations to show that the halo mass function inside cosmic voids is not universal but varies with distance from the void center, with inner regions exhibiting a lower halo number density. Moreover, the measurements of \cite{Verza2022} show that the ratio $\delta_{\rm h}(r) / \delta_{\rm m}(r)$ in concentric spherical shells within voids tends to increase with void-centric distance up to the edge of the void, contradicting the simple constant-bias assumption. The characterization of galaxy bias inside voids, however, remains limited and would certainly benefit from new analytical techniques. 

In this work, we contribute to the understanding of galaxy bias in underdense regions through an analysis of the connection between large-scale linear bias and the position of galaxies inside voids, the latter using a spherical-void identification scheme \citep{Ruiz2015}. This represents a fundamentally different measurement of bias compared to \cite{Verza2022}, which is based on small-scale overdensity ratios. The novelty of our approach lies in two aspects. First, we use a hydrodynamical simulation (IllustrisTNG\footnote{\url{http://www.tng-project.org}}) to model galaxies, rather than relying on simple prescriptions based on halo occupation distributions. Second, we employ an object-by-object estimate of effective large-scale galaxy bias, offering significant analytical advantages compared to the standard method based on ratios of correlation functions or power spectra \citep{Paranjape2018, Contreras2021a, Balaguera2024A, Balaguera2024B, MonteroDorta2025, GarciaFarieta2025}. As described in \cite{Paranjape2018}, the individual bias captures the contribution of each object (in this case, each galaxy) to the sample mean, which, for a given population, coincides with the measurement obtained using the standard approach. Here, we leverage the power of the galaxy-by-galaxy method to evaluate, for the first time, the dependence of bias on a variety of void and tracer properties.

This paper is organized as follows. Section \ref{sec:data} describes the IllustrisTNG data used in this work. The void identification procedure and classification scheme are outlined in Sect.~\ref{sec:voids}, where we also present our fiducial void sample. The method for computing the galaxy-by-galaxy individual bias is described in Sect.~\ref{sec:bias}. The galaxy bias profile inside voids for the fiducial sample is presented and discussed in Sect.~\ref{sec:profile}. The dependence of our results on the tracer sample and identification parameters is addressed in Sect.~\ref{sec:robustness}. Section \ref{sec:interpretation} provides an interpretation of our main results, which are summarized in Sect.~\ref{sec:summary}. The IllustrisTNG300 simulation adopts the standard $\Lambda$CDM cosmology \citep{Planck2016}, with parameters $\Omega_{\rm m} = 0.3089$, $\Omega_{\rm b} = 0.0486$, $\Omega_\Lambda = 0.6911$, $H_0 = 100 \, h\, {\rm km \, s^{-1}Mpc^{-1}}$ with $h=0.6774$, $\sigma_8 = 0.8159$, and $n_s = 0.9667$.

\section{Data}
\label{sec:data}

We use data from the IllustrisTNG magnetohydrodynamical cosmological simulation suite (hereafter referred to as TNG; \citealt{Pillepich2018b,Pillepich2018,Nelson2018_ColorBim,Nelson2019,Marinacci2018,Naiman2018,Springel2018}). These simulations were performed using the {\sc arepo} moving-mesh code \citep{Springel2010} and represent an enhanced version of the earlier Illustris simulations \citep{Vogelsberger2014a, Vogelsberger2014b, Genel2014}. The TNG framework incorporates updated sub-grid physics models, including mechanisms for star formation, radiative metal cooling, chemical enrichment from SNII, SNIa, and AGB stars, as well as feedback processes from stellar populations and supermassive black holes. The simulations were calibrated to reproduce a set of observational constraints, such as the $z=0$ galaxy stellar mass function, cosmic star formation rate density, halo gas fractions, galaxy size distributions, and the black hole--galaxy mass scaling relation. For more details, we direct readers to the references listed above.

\begin{figure}
	\includegraphics[width=\columnwidth]{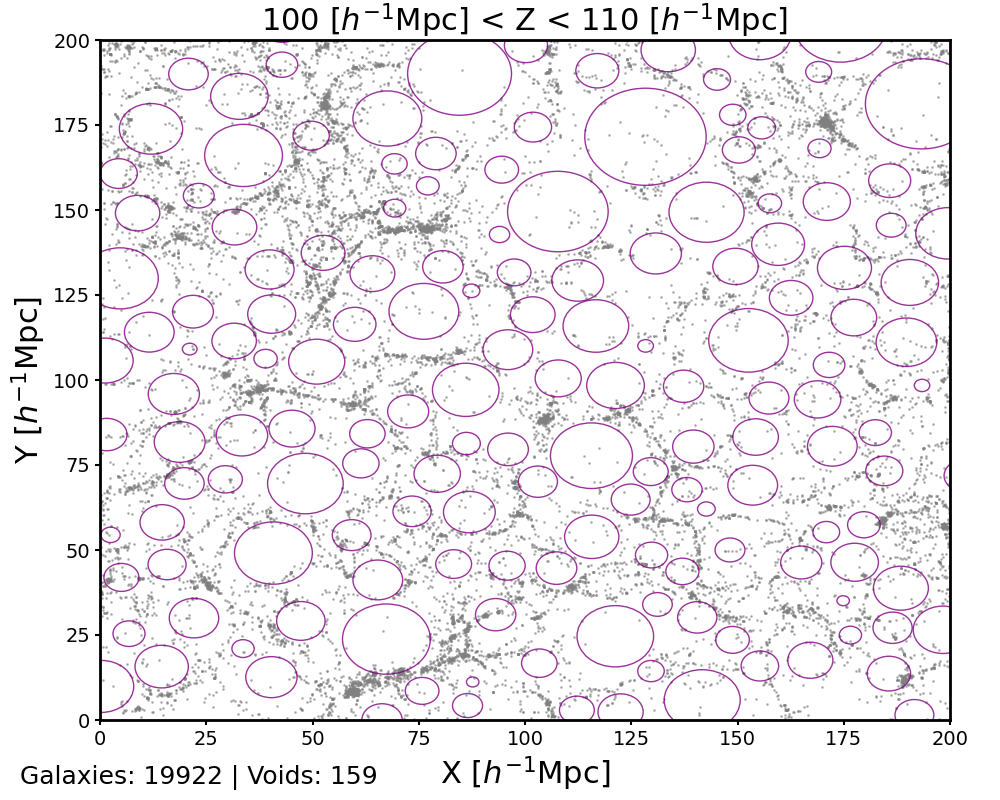}
    \centering
    \caption{ 
    The spatial distribution of galaxies and cosmic voids in a 10-$h^{-1}$Mpc slice ($100 < z[h^{-1}$Mpc$] < 110$) extracted from the TNG300 simulation box. Gray dots represent galaxies in the parent catalog within this region (see Sect. \ref{sec:data}), while purple circles indicate the cross-sectional radii of the corresponding voids from the fiducial sample that intersect this slice (see Sect. \ref{sec:voids}).
}
    \label{fig:scatter+voids}
\end{figure}

\begin{figure*}
    \centering
	\includegraphics[width=1.7\columnwidth]{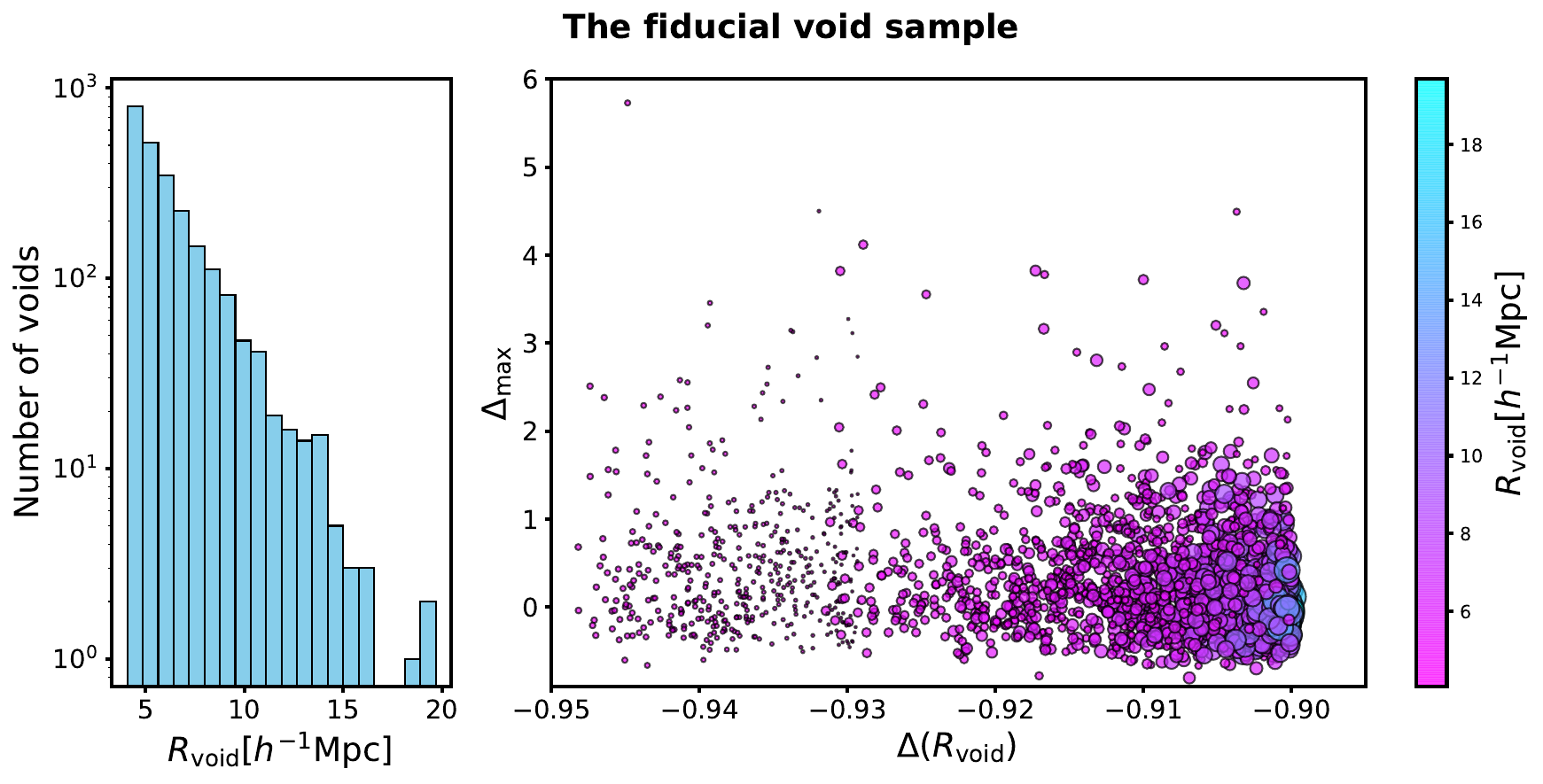}
    \caption{The fiducial void sample extracted from TNG300, based on the selection described in Section \ref{sec:voids}. \emph{Left}: Distribution of void radii in the sample. \emph{Right}: Relation between the overdensity parameters $\Delta(R_{\rm void})$ and $\Delta_{\rm max}$, where symbols are color-coded by void radius, and their sizes are also proportional to void radius.}
    \label{fig:sample}
\end{figure*}

Our analysis focuses on the $z=0$ snapshot of TNG300-1 (hereafter TNG300\footnote{\url{https://www.tng-project.org/data/docs/specifications/}} for simplicity), the largest simulation box in the TNG database. The size of this box, with a volume of 205$^3$ $h^{-3}$Mpc$^3$, and its periodic boundary conditions make it well-suited for LSS analyses. TNG300 follows the evolution of $2500^3$ dark matter particles with a mass of $4.0 \times 10^7 \,\, h^{-1} {\rm M_{\odot}}$ and an initial gas cell resolution of $2500^3$ cells, each with a mass of $7.6 \times 10^6 \,\, h^{-1} {\rm M_{\odot}}$. This dataset has been extensively used to investigate galaxy formation and the interplay between galaxies and their dark-matter halos, contributing valuable insights to a wide range of scientific topics (see \url{https://www.tng-project.org/results/} for a complete list of publications).

Halos in TNG are identified using a friends-of-friends (FOF) algorithm with a linking length set to 0.2 times the mean inter-particle separation \citep{Davis1985}. Subhalos, on the other hand, are detected with the {\sc subfind} algorithm \citep{Springel2001,Dolag2009}, which defines galaxies as subhalos containing a stellar mass component.

Several subhalo and halo properties from TNG are employed in this work. The virial mass of the host halo, $M_{\rm vir}$ [$h^{-1} {\rm M_{\odot}}$], is a key quantity, defined as the total mass enclosed within a sphere of radius $R_{\rm vir}$, where the average density is 200 times the critical density. For galaxies, we primarily consider the stellar mass, $M_\ast{}$ [$h^{-1} {\rm M_{\odot}}$], computed as the sum of the mass of all stellar particles and gas cells bound to each subhalo, along with the 3D Cartesian coordinates and velocities. 

To create our base catalog, a minimal stellar mass cut of $\log_{10}( h^{-1}M_{\rm *}[{\rm M_\odot}]) = 8.33$ is imposed on TNG300. This threshold corresponds to approximately 30 stellar particles. Since we are not classifying galaxies based on their internal properties but are primarily analyzing their spatial distribution, we adopt a fairly inclusive selection\footnote{We have verified, however, that our results are not particularly sensitive to small variations in the minimum mass of the parent catalog.}. This tracer catalog is used to identify our fiducial set of cosmic voids, which are analyzed in Sect. \ref{sec:profile}. The dependence of our results on variations in the minimum stellar mass of tracers is discussed in Sect. \ref{sec:robustness}.

\section{Void identification procedure and classification}
\label{sec:voids}

In this work, cosmic voids are identified using the {\sc Sparkling} public algorithm\footnote{\url{https://gitlab.com/andresruiz/Sparkling}} \citep{Ruiz2015,Ruiz2019}.
In brief, the procedure starts with the estimation of the density field via a Voronoi tessellation of the galaxy maps, which serve as tracers. For each Voronoi cell, the density is computed as $\rho_{\rm cell} = 1/V_{\rm cell}$, with $V_{\rm cell}$ being the volume of the cell. The density contrast is defined as $\delta_{\rm cell}+1=\rho_{\rm cell}/\bar{\rho}$, where $\bar{\rho}$ is the mean density of the tracers. By construction, all Voronoi cells that satisfy $\delta_{\rm cell} < 0$ are identified as the center of an underdense region. For these centers, all spherical volumes with an integrated density contrast $\Delta(R_{\rm void}) \le \Delta_{\rm lim }$ are considered as void candidates, where $R_{\rm void}$ is the void radius and $\Delta_{\rm lim}$ is the density contrast threshold (usually taken as $-0.8$ or $-0.9$, corresponding to 20\% or 10\% the mean density of tracers, respectively). For each of these candidates, the computation of $\Delta$ is repeated several times in randomly displaced centers near the initial locations, sampling the volume of the Voronoi cell's corresponding sphere, updating the center only if the new void radius exceeds the previous one. 
This random sampling procedure is performed to obtain void candidates centered as close as possible to the true local minima of the density field. The final step in the identification process removes overlapping void candidates, retaining only the largest voids that do not overlap with others.

To construct our fiducial void catalog, we use the following criteria. First, we identify voids using the entire parent catalog as tracers. Second, we set the maximum integrated overdensity within $R_{\rm void}$, $\Delta(R_{\rm void})$, to $\Delta_{\rm lim} = -0.9$. Note that this value is motivated by the spherical evolution model for the dark-matter density field \cite[e.g.][]{Sheth_voids}, but this same value is often used for biased tracers as well. Finally, we conduct the fiducial analysis in real space. Figure \ref{fig:scatter+voids} shows, as an example, the fiducial voids intersecting a slice of 10 $h^{-1}$Mpc (100 < $z$[$h^{-1}$Mpc] < 110) from the TNG300 box. Figure \ref{fig:sample} further illustrates key characteristics of this sample, including the distribution of void sizes (left) and the relationship between two important overdensity properties: the aforementioned $\Delta(R_{\rm void})$ and the maximum overdensity in the surrounding environment, $\Delta_{\rm max}$. We define this quantity as the maximum overdensity between $[2R_{\rm void}, 3R_{\rm void}]$, which is closely linked to the dynamical evolution of voids, as discussed below. Even though we set $\Delta_{\rm lim} = -0.9$, Fig. \ref{fig:sample} shows that $\Delta(R_{\rm void})$ takes values in the range [-0.95, -0.9]. Note that, by construction, voids are identified counting distance-sorted tracers from the void center. In this scheme, the discrete nature of the tracers plays an important role, particularly when spherical volumes and integrated densities are computed in low-density environments. Discreteness adds shot-noise to the overdensity parameter at the void radius, such that the larger the number density of tracers, the closer the value of $\Delta(R_{\rm void})$ to $\Delta_{\rm lim}$. This shot-noise also explains why the smallest voids do not necessarily have the smallest values of $\Delta(R_{\rm void})$ (see the artificial discontinuity between -0.94 and -0.93 in Fig. \ref{fig:sample}).

Our procedure yields a total of 2,393 voids for the fiducial sample. Adopting, again, an inclusive selection is motivated by the goal of optimizing the statistics in the measurement of the bias profiles. In following sections, we will analyze this sample and investigate how variations in the tracer population and $\Delta_{\rm lim}$ impact our results.

As part of the void identification scheme, each void is assigned a type based on their surrounding large-scale environment, following \cite{Ceccarelli_clues_2013}. According to this classification, R-type voids are surrounded by large-scale underdense regions, whereas S-type voids are embedded in overdense regions. As shown in \cite{Paz_clues_2013}, regions surrounding R-type voids are naturally expanding, whereas those around S-type voids are contracting. We use $\Delta_{\rm max}$ to impose this distinction. A void is considered S-type if $\Delta_{\rm max}\ge 0$, while $\Delta_{\rm max}<0$ corresponds to R-type voids.

\begin{figure*}
	\includegraphics[width=1.9\columnwidth]{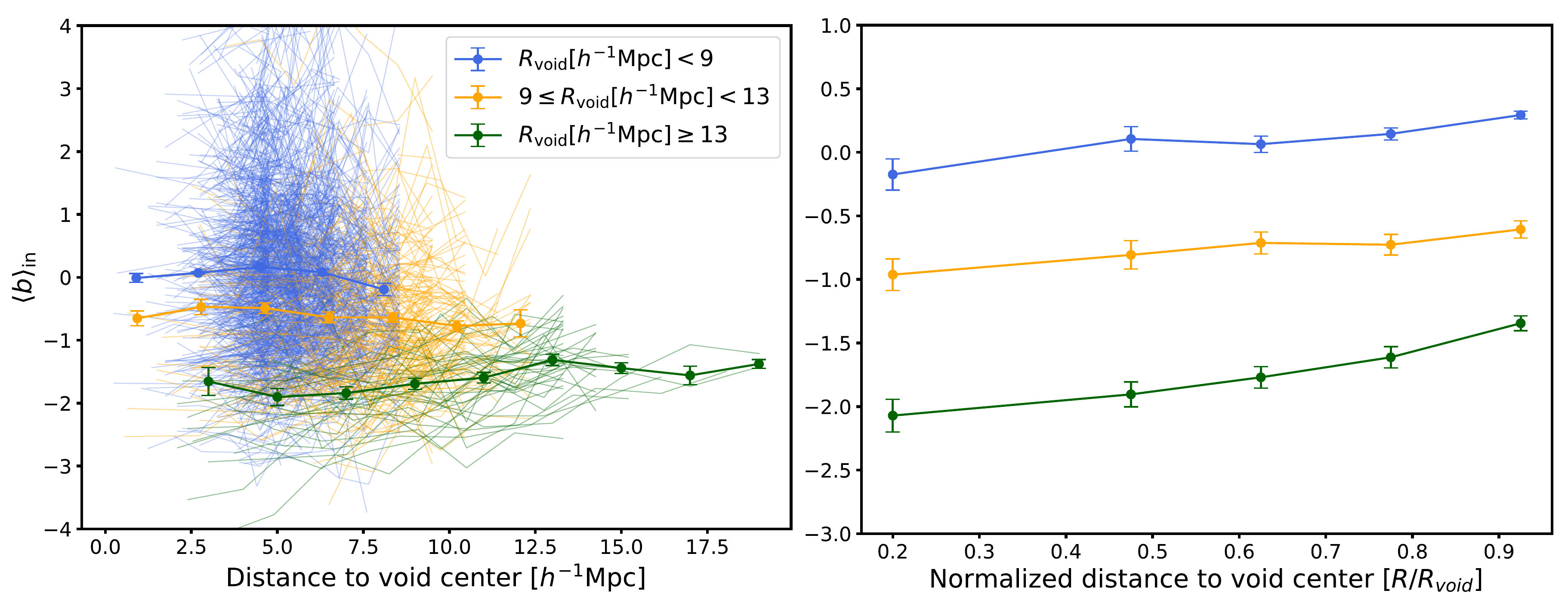}
    \centering
    \caption{The galaxy bias profile inside voids as a function of void size. \emph{Left}: The average galaxy bias as a function of distance to the center of each void in our fiducial void sample. Three different void radius ranges are indicated by color. The average galaxy bias profile for each of these subsets are also shown, where uncertainties correspond to the errors on the means. \emph{Right}: The average galaxy bias profile for the same  subsets but using the void-centric distance normalized by void radius. Only central galaxies are employed in this analysis. A non-uniform binning has been adopted to avoid discreteness issues in the innermost regions of voids (see text). This binning is maintained in all figures.}
    \label{fig:bias_profile_all}
\end{figure*}

\begin{figure}
	\includegraphics[width=1\columnwidth]{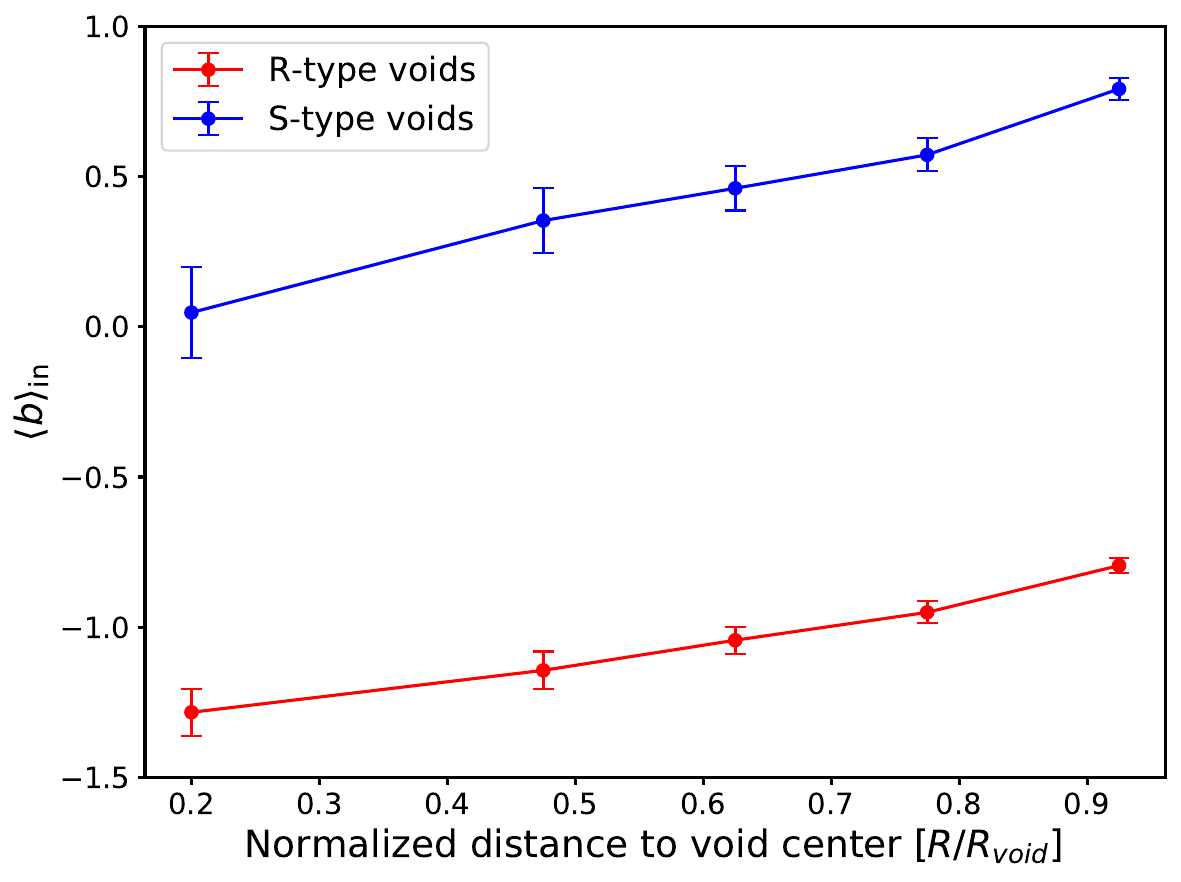}
    \caption{The average galaxy bias profile for R-type and S-type voids, where the distance to the center of voids have been normalized by the void radius, $R_{\rm void}$. The uncertainties correspond to the errors on the means.}
    \label{fig:profile_types}
\end{figure}

\section{Assigning individual galaxy bias}
\label{sec:bias}

The standard way of measuring the large-scale (linear) halo bias is based on a ratio of either correlation functions or power spectra, e.g., $b = \xi_{hm}/\xi_{mm}$, where $\xi_{mm}$ denotes the auto-correlation of the dark-matter density field, and $\xi_{hm}$ represents the cross-correlation between halos and the dark matter. A drawback of these methods is that the bias is measured for subsets of halos, potentially complicating analyses that seek to disentangle correlations among internal halo properties, environment, and bias. Instead, it is sometimes advantageous to use an object-by-object estimate of halo bias, as developed by \cite{Paranjape2018}. In this framework, the effective large-scale bias is viewed as a sample mean, where each element contributes with an individual bias.  This individual-bias approach has been applied in several studies (e.g., \citealt{Ramakrishnan2019, Contreras2021a, Balaguera2024A, Balaguera2024B, MonteroDorta2025}). These works have shown that the mean of the individual halo bias in bins of halo mass (i.e., the so-called \emph{bias function}) is in good agreement not only with measurements with the traditional method \citep[see e.g.][]{Tinker2010,2012MNRAS.420.3469P}, but also with analytical derivations based on the peak height of density fluctuations (e.g., \citealt{Mo1996}).  

Here, we apply the philosophy of \cite{Paranjape2018} to compute the large-scale linear bias of each galaxy in our sample. Following the prescription of \cite{Balaguera2024A}, the individual linear galaxy bias of a galaxy $i$ at position $\mathbf{r}$, $b_i$, is given by  
\begin{equation}
    b_i =\frac{\sum_{j,k_{j}<k_{max}}N^{j}_{k}\langle \exp[-i\bf{k}\cdot \bf{r}_{i}]\delta_{\mathrm{DM}}^{*}(\bf{k}) \rangle_{k_{j}}}{\sum_{j,k_{j}<k_{max}} N^{j}_{k}P_{\rm DM}(k_{j})},
   \label{eq:bias}
\end{equation}
where $\delta_{\mathrm{DM}}(\mathbf{k})$ is the Fourier transform of the dark-matter density field, $P_{\rm DM}(k_{j})$ is the matter power spectrum, and $N^{j}_{k}$ is the number of Fourier modes in the $j$-th spherical shell\footnote{The assignment of individual galaxy bias has been performed using the \texttt{CosmiCCcodes} library at \url{https://github.com/balaguera/CosmicCodes}}. The sum runs over the range of wavenumbers in which the ratio between the galaxy and dark-matter power spectra remains constant. Given the volume of the simulation, we use the range $k \leq 0.2 \,h \, \rm{Mpc}^{-1}$, up to which the ratio between the galaxy and dark-matter power spectra is compatible with a constant value, as we have verified. In this framework, the effective large-scale bias of a given population of $N_{\rm G}$ galaxies can be trivially computed as the mean of all the individual biases of the galaxies in the subset, namely:

\begin{equation}
    \langle b \rangle_{\rm G} = \frac{\sum_{i = 1, N_{\rm G}} b_i}{N_{\rm G}},
   \label{eq:bias_mean}
\end{equation}

It is important to bear in mind that, rather than describing a local connection between galaxies and dark matter, Eq.~(\ref{eq:bias}) represents the contribution of each galaxy to the bias of the population on large scales (Eq. \ref{eq:bias_mean}).

\begin{figure*}
	\includegraphics[width=2\columnwidth]{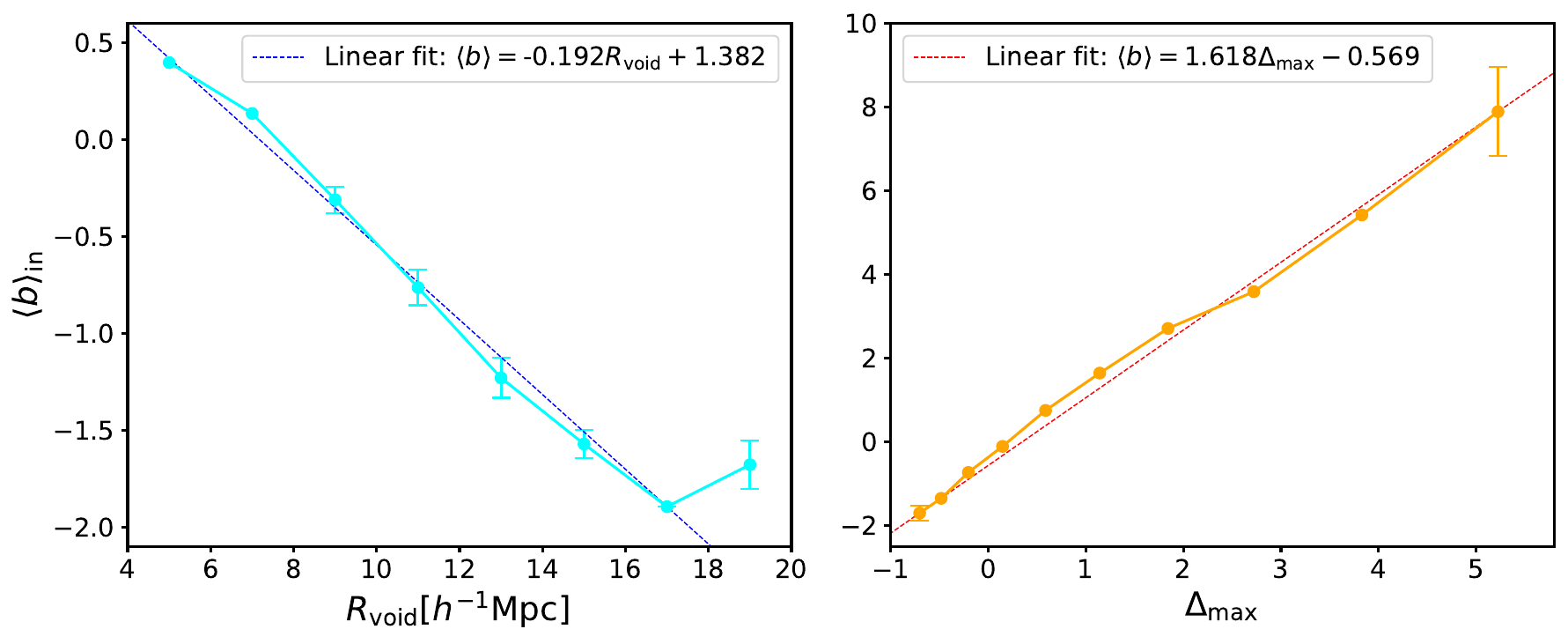}
    \caption{The average galaxy bias as a function of void radius, $R_{\rm void}$ (\emph{left}), and maximum overdensity in the surrounding region, $\Delta_{\rm max}$ (\emph{right}). The uncertainties correspond to the errors on the means. Note that, as shown in Fig. \ref{fig:sample}, voids with high values of $\Delta_{\rm max}$, which also tend to have high bias, are very scarce. At fixed $R_{\rm void}$, the population is dominated by lower-$\Delta_{\rm max}$ (lower-$\langle b \rangle$ voids), which explains why the average bias only reaches $\sim$0.5 in the left panel of this figure.}
    \label{fig:average_bias}
\end{figure*}

\begin{figure}
	\includegraphics[width=1\columnwidth]{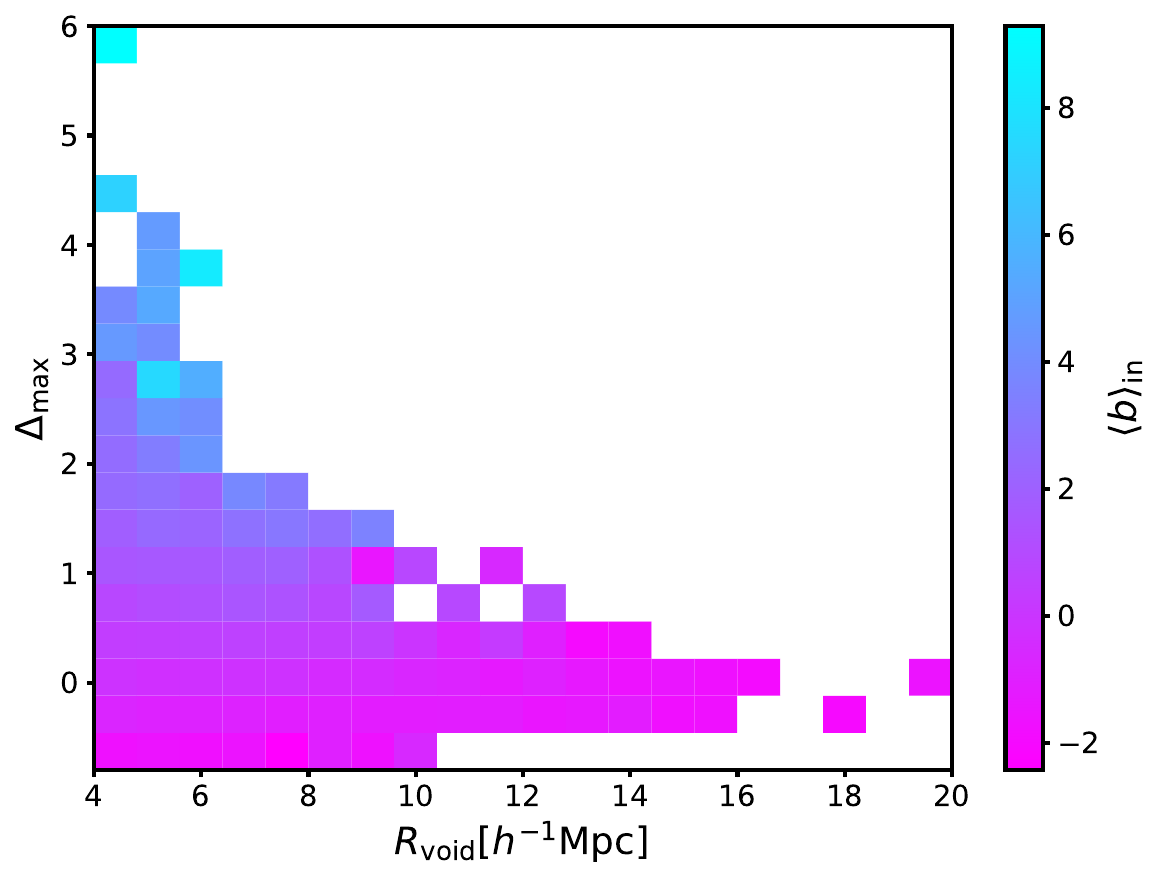}
    \caption{A 2D color map illustrating how galaxy bias varies as a function of both void radius ($R_{\rm void}$) and maximum surrounding overdensity ($\Delta_{\rm max}$). The color code indicates the mean galaxy bias inside the voids pertaining to each pixel.}
    \label{fig:colormap}
\end{figure}

\begin{figure*}
    \includegraphics[width=2\columnwidth]{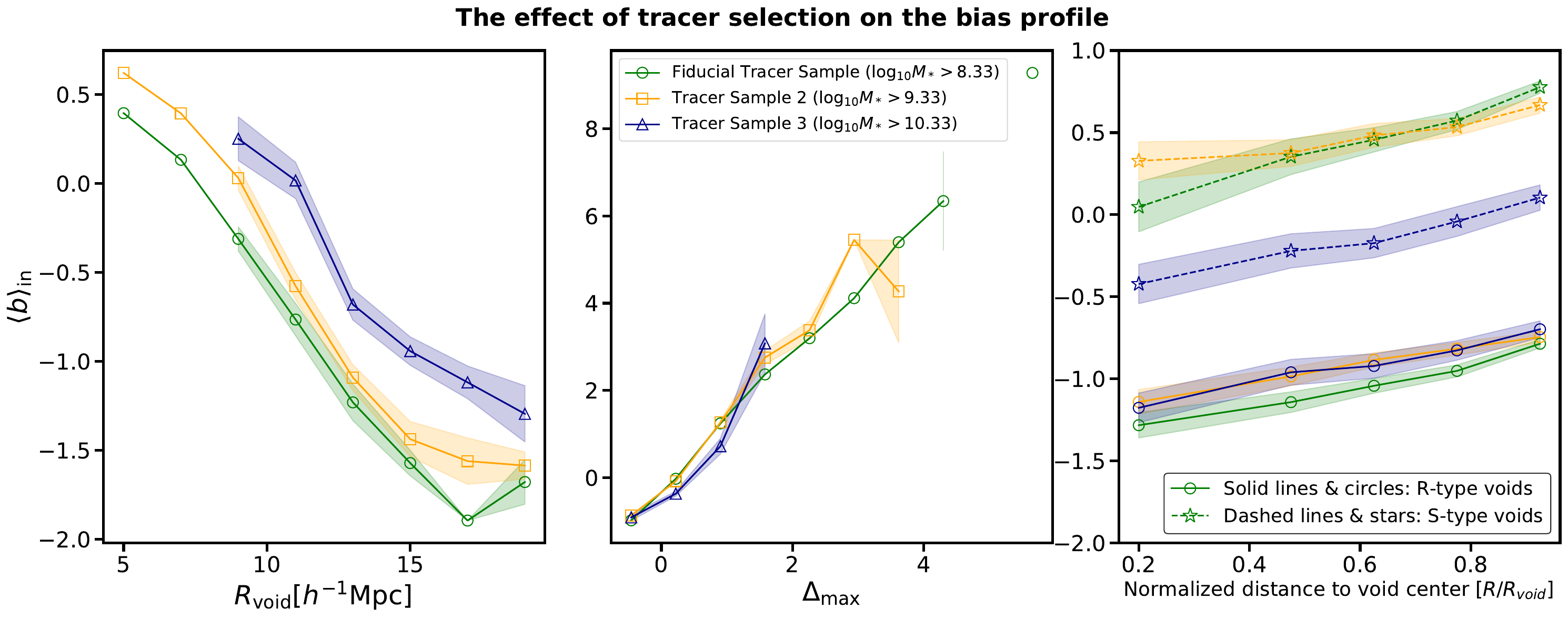}
    \caption{The bias--size and bias--maximum surrounding overdensity relations (left, middle), along with the dependence of the bias profile on void type (right), for three different tracer samples with varying minimum stellar mass. The results shown in green correspond to our fiducial void sample, obtained from a galaxy catalog with $\log_{10}( M_{\rm *}[h^{-1}{\rm M_\odot}]) > 8.33$ (same as Figs. \ref{fig:profile_types} and \ref{fig:average_bias}). Yellow and blue lines/symbols show the effect of increasing this threshold to $\log_{10}( M_{\rm *}[h^{-1}{\rm M_\odot}]) > 9.33$ and 10.33, respectively. Uncertainties (shaded regions) correspond again to the errors on the means.}
    \label{fig:mass_dependence}
\end{figure*}

\begin{figure*}
    \includegraphics[width=2\columnwidth]{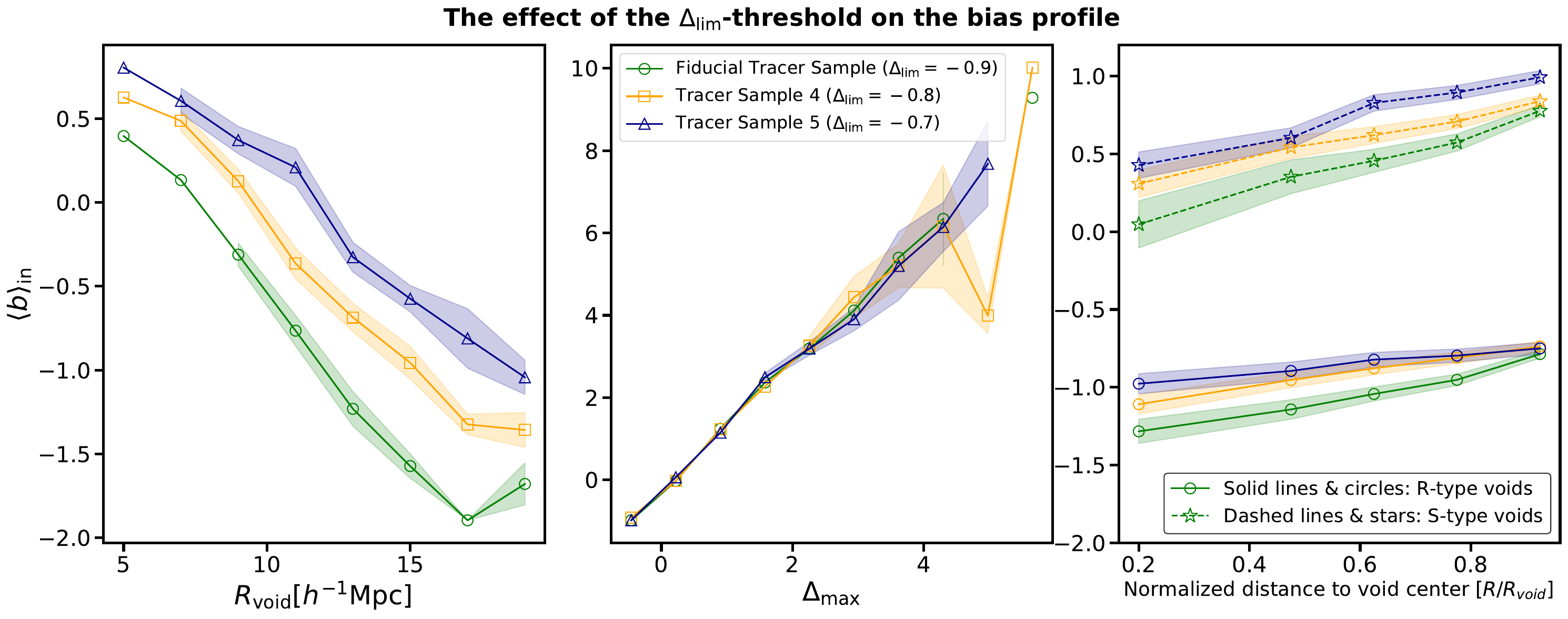}
    \caption{Same as Fig. \ref{fig:mass_dependence} but varying the overdensity threshold, $\Delta_{\rm lim}$ -- one of the main parameters used to define voids. The green lines/symbols show again the fiducial sample (with $\Delta_{\rm lim} = -0.9$), whereas results for $\Delta_{\rm lim} = -0.8$ and $-0.7$ are represented in yellow and blue, respectively.}
    \label{fig:density_dependence}
\end{figure*}

\begin{figure*}
    \includegraphics[width=2\columnwidth]{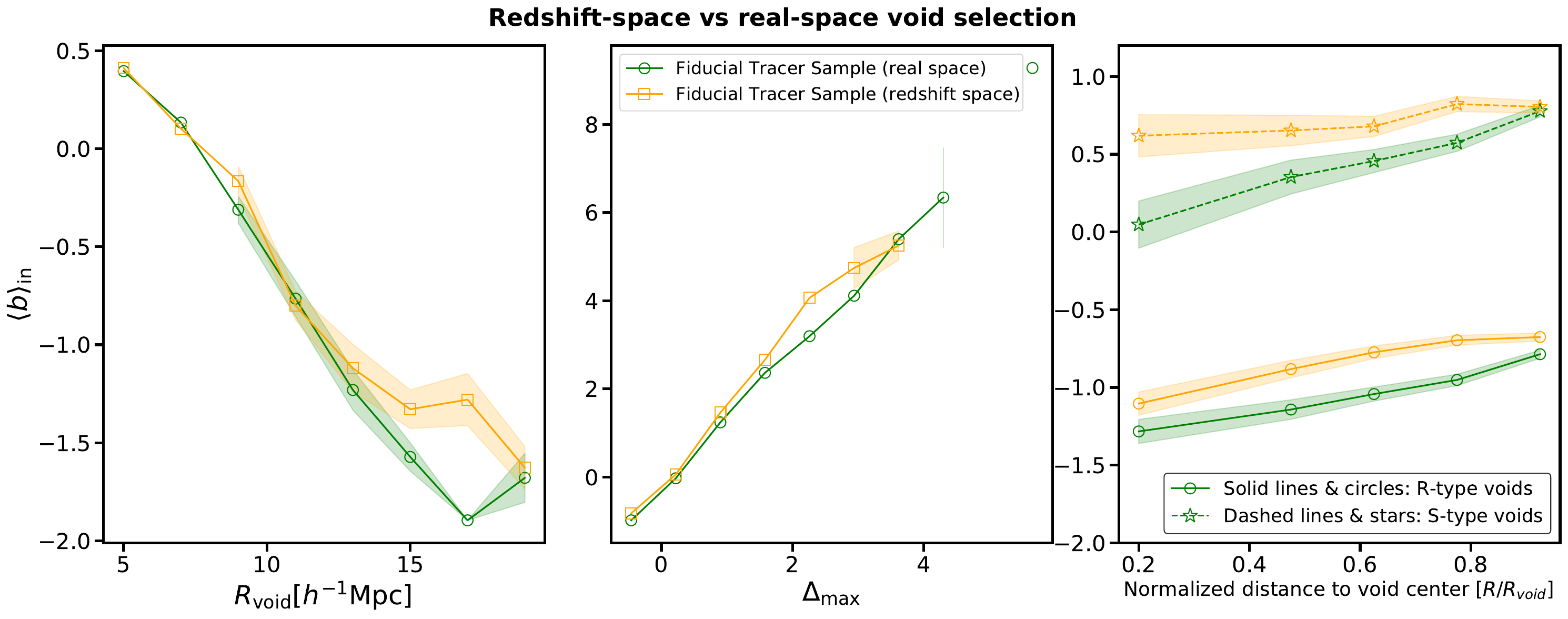}
    \caption{The same as Fig. \ref{fig:mass_dependence} but focusing on the impact of a void identification scheme performed in redshift space. The green lines/symbols show the fiducial sample (real-space), whereas results in yellow are obtained by simulating a redshift-space void identification (see text).}
    \label{fig:redshift_space}
\end{figure*}

\section{The void bias profile}
\label{sec:profile}

In this section, we analyze the dependence of the individual contributions to the effective large-scale galaxy bias on the distance to the void center -- what we call here \emph{the galaxy bias profile} -- for our fiducial set of voids. For simplicity, and given our interest in large-scale bias, the computation of the galaxy bias profiles (an any related quantity) is restricted to central galaxies throughout this work. This has, nevertheless, little effect on our results, due to the intrinsic characteristics of the halo populations in voids.

Figure \ref{fig:bias_profile_all} displays these measurements for the entire sample, where voids have been divided into 3 groups based on their radii:  $R_{\rm void}[h^{-1}\rm Mpc]< 9$, $9\le R_{\rm void}[h^{-1}\rm Mpc]< 13$, and $R_{\rm void}[h^{-1}\rm Mpc]\ge 13$. We show both the individual profiles and the average stacked measurements for these three subsets as a function of void-centric distance, both in units of $h^{-1}\rm Mpc$ (left) and normalized to void radius, i.e., $R/R_{\rm void}$ (right). For the latter, a non-uniform binning based on the following edges is imposed to avoid discreteness effects in the innermost regions of the voids, where statistics are poor: [0.0, 0.4, 0.55, 0.7, 0.85, 1.0]\footnote{We have verified that our results remain qualitatively unchanged when a uniform binning is adopted, although the profiles become noisier in the inner regions.}. Three important conclusions can be extracted from this figure. First, the average bias is higher for smaller voids; voids with $R_{\rm void}[h^{-1}\rm Mpc]< 9$ have an average bias of $\sim 0$, whereas the stacked profile for the largest voids yields average biases in the range $[-2,-1]$. Second, while the profiles do not show a clear trend when averaged as a function of distance in physical units (left), a tendency toward increasing bias emerges when the profiles are stacked as a function of distance in the more natural normalized units (right). This trend is particularly significant for the largest voids. In contrast, the stacked profiles for small and intermediate-sized voids tend to flatten at $R \gtrsim 0.5\, R_{\rm void}$. Third, the small-void population appears to be more heterogeneous in bias, as indicated by the larger scatter in the profiles. However, these voids are also the most common objects (see Fig. \ref{fig:sample}).

Another interesting way to divide the sample is by void type. According to the classification by \cite{Ceccarelli_clues_2013}, described in Sect. \ref{sec:voids}, voids can be categorized into two types. S-type voids (those with $\Delta_{\rm max} \geq 0$) are embedded in overdense regions, whereas R-type voids ($\Delta_{\rm max} < 0$) are immersed in low-density regions. A physical model for this difference was discussed in \cite{MassaraSheth2018}. Figure \ref{fig:profile_types} displays the stacked bias profiles as a function of distance, expressed in units of $R_{\rm void}$, for these two void types. The figure shows that, on average, galaxies inside S-type voids have a significantly higher bias than those inside R-type voids. Although the overall profiles appear fairly similar, S-type voids exhibit slightly steeper profiles. These results may serve as the basis for potential extensions of the \cite{MassaraSheth2018} model.

The results shown in Fig. \ref{fig:profile_types} are consistent with Fig. \ref{fig:bias_profile_all}, as S-type voids tend to be smaller than R-type voids. The average radius of an R-type void in our fiducial sample is $6.55 \pm 2.39$ $h^{-1} \rm Mpc$, while it is $5.82 \pm 1.77$ $h^{-1} \rm Mpc$ for S-type voids. Looking back at Fig. \ref{fig:bias_profile_all}, S-type voids dominate the small-radius subset (blue), comprising $65\%$ of it. In contrast, the large-radius subset (green) is R-type dominated ($65\%$), while the intermediate subset (orange) is quite heterogeneous ($54\%$ R-type).  

Figure \ref{fig:bias_profile_all} shows that, on average, galaxy bias inside voids tends to increase with void radius. For R-type voids, this is simply because the center of a void is the most underdense, so it has the most negative large-scale bias, whereas its edges, being less underdense, are less anti-biased. For S-type voids, the void walls are overdense, so there the bias is naturally positive. This relationship becomes clearer when the mean galaxy bias is computed across the entire void (i.e., averaged over all distances).  Figure \ref{fig:average_bias} (left panel) shows the mean large-scale bias, $\langle b\rangle_{\rm in}$, as a function of void size for our fiducial sample. For intermediate-size voids, the relation is approximately linear -- with a slope of $-0.19$ -- but it flattens at both ends, particularly for the largest voids. The right-hand panel shows that the mean galaxy bias also scales strongly with $\Delta_{\rm max}$, the maximum overdensity within $2$ and $3R_{\rm void}$. The average bias shows an approximately linear increase up to $\Delta_{\rm max} \simeq 2$, where the trend breaks. Overall, a linear model with a slope of $\simeq 1.62$ provides a reasonable fit to the data. This correlation is not surprising, since $\Delta_{\rm max}$ is the mean density on a scale that is comparable to $R_{\rm void}$, whereas $\langle b\rangle_{\rm in}$ is related to the larger-scale density field centered on the same void.

Finally, the information presented in Fig. \ref{fig:average_bias} is extended in the bias map of Fig. \ref{fig:colormap}, which illustrates the variation of bias across the $\Delta_{\rm max} - R_{\rm void}$ plane. This figure allows the reader to identify the exact regions of the parameter space where the bias is either maximized or minimized. At fixed void radius, the bias depends strongly on the density profile of the surrounding environment, characterized by $\Delta_{\rm max}$, especially for smaller voids (the steeper the profile, the higher the bias). Figure \ref{fig:colormap} also completes the description of R- and S-type voids, showing that the former display less variation in bias and a weaker correlation between $\Delta_{\rm max}$ and $R_{\rm void}$.

\section{The effect of tracer selection}
\label{sec:robustness}

Our results so far are based on a specific choice of galaxy tracer sample used to identify voids ($\log_{10}( M_{\rm *}[h^{-1}{\rm M_\odot}]) > 8.33$, $\Delta_{\rm lim} = -0.9$, real-space). Some variation in our results is expected, at least quantitatively, when this selection is modified. In this section, we analyze the impact of changing this selection on the bias-radius and bias-$\Delta_{\rm max}$ relations (Fig. \ref{fig:average_bias}), as well as on the dependence on void type (Fig. \ref{fig:profile_types}).  

Figure~\ref{fig:mass_dependence} focuses on the dependence on the stellar mass of the tracers. In each panel, results based on the fiducial sample are shown in green, whereas the yellow and blue symbols/lines represent voids identified with tracer samples with minimum stellar masses of $\log_{10}( M_{\rm *}[h^{-1}{\rm M_\odot}]) = 9.33$ and 10.33, respectively. The effect of this change is evident on the $\langle b \rangle_{\rm in} - R_{\rm void}$ relation, which is significantly shifted upward for more massive tracer samples, particularly the most restrictive one. Increasing the minimum stellar mass for tracers implies that the resulting voids will host objects with higher masses, which tends to increase the average bias for a fixed void radius. Considering only the most massive tracers ($\log_{10}( M_{\rm *}[h^{-1}{\rm M_\odot}]) > 10.33$) also has an impact on void size, as only voids with $R_{\rm void} \gtrsim 9$ $h^{-1}\rm Mpc$ are sufficiently well mapped. 

Despite the significant impact on the $\langle b \rangle_{\rm in} - R_{\rm void}$ relation, the $\langle b \rangle_{\rm in} - \Delta_{\rm max}$ relation is approximately preserved (middle panel). This is nontrivial, since a change in the tracer selection implies that the region that defines the $\Delta_{\rm max}$ value also changes ($\Delta_{\rm max}$ is calculated based on the tracer density field). Nevertheless, at fixed $\Delta_{\rm max}$, the average bias remains fairly unaffected by stellar mass. The dependence on the type of voids (right panel) changes primarily for the most massive tracer sample, in the sense that the bias of S-type voids is lower than in the fiducial and intermediate-mass subsets.  Importantly, the general shape of the profiles remains almost intact, except for some small variations in the inner regions.

Another important parameter in the void identification process is the limit overdensity integrated within one $R_{\rm void}$, $\Delta_{\rm lim}$, which was set to $-0.9$ for our fiducial void sample. In Fig. \ref{fig:density_dependence}, we evaluate the impact of varying this parameter on the same relations shown in Fig. \ref{fig:mass_dependence}. The figure reveals that increasing $\Delta_{\rm lim}$, i.e., making the voids less underdense, has a similar effect to increasing the stellar-mass threshold: the average bias within a void becomes less negative (or more positive). This is consistent with expectations based on the bias of massive galaxies and the natural connection between bias and galaxy overdensity \citep{Mo1996,  ShethTormen2002}. Therefore, the dependence of the scaling relations on $\Delta_{\rm lim}$ appears more linear and smoother than that of stellar mass (Fig. \ref{fig:mass_dependence}). The $\langle b \rangle_{\rm in} - \Delta_{\rm max}$ relation (middle panel) is nearly independent of $\Delta_{\rm lim}$. This is because variations in $\Delta_{\rm lim}$ have little effect on the position of the void center or on $\Delta_{\rm max}$, as they do not significantly alter the tracer density around voids. Regarding the dependence on void type, increasing $\Delta_{\rm lim}$ produces a global upward shift in all trends, further illustrating the strong connection between overdensity and bias.  

Finally, we have studied the impact of identifying voids in redshift space instead of real space on the bias relations. To simulate a redshift-space void identification, we displaced the positions of galaxies in the $z$-direction in our parent sample based on the corresponding component of the subhalo velocity provided by TNG300. Voids were subsequently identified assuming an overdensity limit of $\Delta_{\rm lim} = -0.9$, to mimic the fiducial selection. The comparison between the resulting and fiducial void samples is presented in Fig. \ref{fig:redshift_space}. In terms of the bias-size relation, a significant difference arises only for the largest voids, with those having $R_{\rm void}[h^{-1} \rm Mpc] \gtrsim 15$ showing, on average, an enhanced bias. This bias enhancement is also evident in the bias-$\Delta_{\rm max}$ relation and the R-type and S-type bias profiles, all of which are shifted upward, particularly the bias profiles. We note that this is a simplified exercise to explore the effects associated with an observational study. A more realistic analysis would include a bias measurement performed directly in redshift space.  

\begin{figure*}
	\includegraphics[width=2\columnwidth]{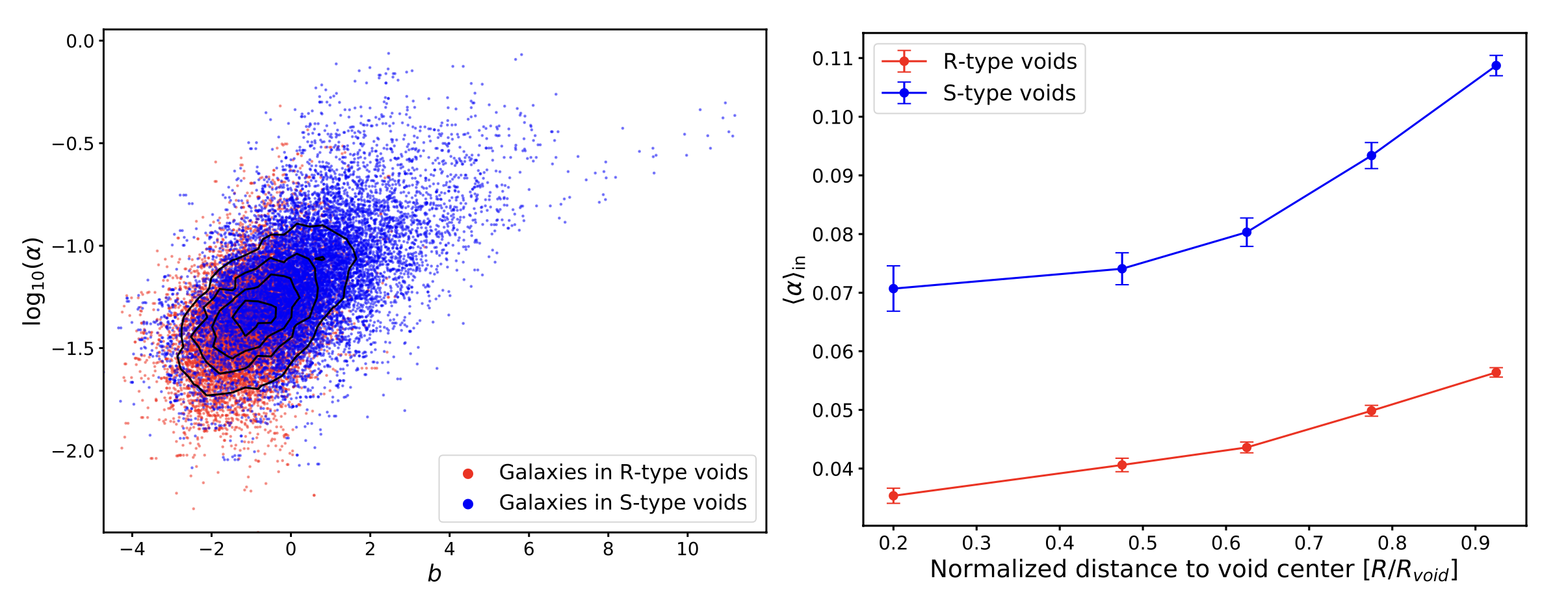}
    \caption{The correlation between individual large-scale bias, $b$, and local (small-scale) tidal anisotropy, $\alpha$, inside voids. \emph{Left}: Correlation for galaxies inside S- and R-type voids, separately (contours representing the combined population are added for reference). \emph{Right}: The $\alpha$ profile (i.e., the average $\alpha$ as a function of normalized void-centric distance) for these two void types. Uncertainties correspond to the errors on the means.}
    \label{fig:calpha}
\end{figure*}

\begin{figure*}
    \includegraphics[width=2\columnwidth]{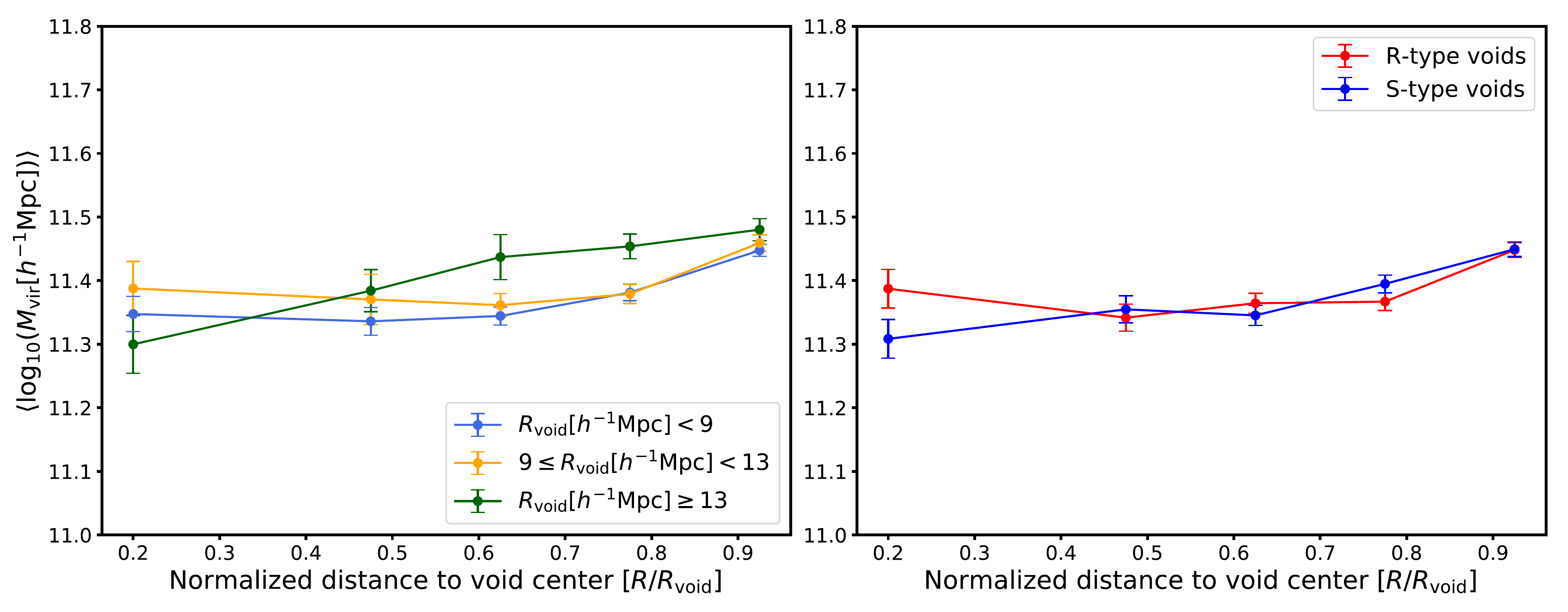}
    \caption{The halo mass profile as a function of void size (left) and  type (right), following the same formats of Figs. \ref{fig:bias_profile_all} and \ref{fig:profile_types}. Uncertainties correspond to the errors on the means. No significant differences in the mean halo mass as a function of normalized distance are observed for the subsets under consideration.}
    \label{fig:conclusions}
\end{figure*}

\section{Interpretation \& Discussion}
\label{sec:interpretation}

Cosmic voids are intriguing regions within the cosmic web. They occupy a significant fraction of the Universe's volume while featuring the lowest matter densities \citep{Sheth_voids}. These low densities result in unique properties for the galaxies and halos that form inside them, separating  them from those in other cosmic environments. Due to the intrinsically low number densities of galaxies within voids, characterizing the connection between galaxies and the underlying matter density field -- described by the bias -- has been particularly challenging. In this work, we use an object-by-object estimate of large-scale linear galaxy bias to perform a detailed analysis of galaxy bias within voids, examining its connection to the global properties of voids and their surrounding environments. 

As expected, the large-scale effective bias of galaxies inside voids is, on average, very low, often reaching values well below zero (remember, this is the average contribution of each galaxy to the large-scale bias of the population). This result is qualitatively consistent with measurements of individual large-scale halo bias inside voids in the UNIT simulation, as reported in \cite{Balaguera2024A}. In that work, the cosmic web is characterized using the eigenvalues of the tidal tensor, with the average halo bias inside voids reaching values of around $-2$, similar to those of our largest voids (see their figure 16). It is important to emphasize that this result should be interpreted on large scales, rather than as a local relation between overdensities: the population of galaxies inside voids is significantly less clustered than the matter density field on scales larger than the one corresponding to the minimum $k$ in Eq.~(\ref{eq:bias}). Our findings are also consistent with the measurement of relative bias as a function of the distance to saddles and nodes at fixed halo mass presented in \cite{MonteroRodriguez2024}, where the standard method based on ratios of correlation functions was employed. In that analysis, the relative bias of central galaxies is computed with respect to the entire population in the same halo mass bin, over scales of 5-15 $h^{-1} \rm Mpc$. As shown in their figure 3, the relative bias becomes negative for galaxies located far from saddles and nodes, where the correlation function itself turns negative for this population.

We have also shown that the linear large-scale bias computed inside voids changes as a function of the distance to the void center when the distance is normalized by the void radius. The stacked galaxy bias profile inside voids increases outward as we move closer to the void walls. This result is, again, consistent with measurements of bias as a function of the distance to the critical points of the cosmic web presented in \cite{MonteroRodriguez2024}. Our findings are also related to those of \cite{Verza2022}, albeit not directly comparable. In that work, the halo mass function is shown to depend on void-centric distance, from which a trend of increasing halo bias as a function of distance is derived. Recall that the bias in that case is defined as the relation between halo and matter overdensities on concentric spherical shells, rather than as a large-scale bias.

A crucial feature that influences the properties of voids is the surrounding environment, typically characterized by the maximum overdensity around voids, $\Delta_{\rm max}$. This parameter is used to define two classes of voids \citep{Ceccarelli_clues_2013}: R-type voids are surrounded by large-scale underdense regions, whereas S-type voids are embedded in overdense regions. As shown in \cite{Paz_clues_2013}, regions surrounding R-type voids are naturally expanding, while those around S-type voids are contracting. Halos within R-type voids also tend to experience slower growth compared to those in S-type voids \citep{RodriguezMedrano2022}. Furthermore, galaxies in R-type voids are both bluer and more centrally concentrated \citep{RodriguezMedrano2022}. Our measurements add to this characterization by revealing a notable difference in large-scale bias between these two types of voids, with galaxies inside R-type voids exhibiting significantly lower bias than those in S-type voids. This result suggests that bias follows the galaxy distribution, as voids with positive $\Delta_{\rm max}$ (S-type) reach higher galaxy densities. Additionally, there is a clear connection between $\Delta_{\rm max}$ and void size, with larger voids having lower $\Delta_{\rm max}$, typically converging to a value around zero. This relationship explains the strong correlation between void radius and galaxy bias, where galaxy bias decreases with increasing void size.

The connection between linear bias and the cosmic web can also be characterized using the tidal anisotropy parameter, $\alpha$,  which is built from the eigenvalues, $\lambda_i$, of the tidal field tensor\footnote{The elements of the tidal tensor are defined as $\mathcal{T}_{ij}=\partial_{i}\partial_{j}\phi$, where $\phi$ is the comoving gravitational potential satisfying the Poisson equation $\nabla^{2}\phi=\delta_{dm}$, with $\delta_{dm}$ being the dark-matter overdensity.}, as mentioned before. Inspired by \cite{Paranjape2018} and following the definition of \cite{Balaguera2024A}, the tidal anisotropy can be computed as
\begin{equation}
\alpha \equiv \frac{\sqrt{(\lambda_{1}-\lambda_{2})^{2}+(\lambda_{1}-\lambda_{3})^{2}+(\lambda_{3}-\lambda_{2})^{2}}}{2(2+\lambda_{1}+\lambda_{2}+\lambda_{3})}
\label{eq:alpha}
\end{equation}

Halos in more anisotropic local environments (higher $\alpha$) are more clustered on large scales than halos of similar mass in more isotropic environments, such as voids \citep{Paranjape2018, Alam2019, Balaguera2024A}. The voids we are studying are defined as underdense spherical regions centered around a density minimum. On average, these regions contain a homogeneous distribution of objects in their innermost areas, while the density gradually increases toward the walls \citep[see][]{Ceccarelli_clues_2013, Paz_clues_2013}. Naturally, the anisotropy is expected to be greater in the outer regions, as we move closer to the filaments \cite[see][for illustrative toy models]{Paranjape2021}. 

In order to explore this connection, we have computed $\alpha$ for all galaxies in our parent catalog. Importantly, the eigenvalues in Eq. \ref{eq:alpha} depend on smoothing scale (e.g., for halos, $\alpha$ on the turnaround scale shows the strongest correlations with internal halo properties and their large-scale environments, see \citealt{Paranjape2018, Ramakrishnan2019}). Here, we simply use a fixed scale of $\sim 2h^{-1}$Mpc -- approximately the turnaround scale of the halos used to identify the voids, but not necessarily the optimal scale for the voids themselves -- and leave a more detailed exploration of scale to future work. The left panel of Fig. \ref{fig:calpha}, where we present these results, displays a clear correlation between this local (i.e. small-scale) $\alpha$ and large-scale bias for central galaxies inside R-type and S-type voids, separately. This figure also demonstrates that the anisotropy inside S-type voids is on average higher than inside R-type voids. The correlation between $\alpha$ and $b$ also dictates the distinct shapes of the $\alpha$-profiles of R-type and S-type voids, which follow qualitatively the bias profiles\footnote{One should resist the temptation to treat these $\alpha$-profiles as the direct analogs of the large-scale bias profiles, since $\alpha$ and $b$ have been computed over different scales (i.e., $\alpha$ includes here small-scale information). We are in the process of generalizing Eq.~1 to estimate the object-by-object tidal shear.} (right panel of Fig. \ref{fig:calpha}). 

Importantly, the bias trends reported in this work are not driven by halo or stellar mass -- that is, they are not the result of distinct mass distributions across the different samples. To demonstrate this, Fig. \ref{fig:conclusions} presents the average host halo mass profile (i.e., the average halo mass as a function of distance from the void center normalized by the void radius) for subsets of voids based on size (left panel, same samples as in Fig. \ref{fig:bias_profile_all}) and type (right panel, as in Fig. \ref{fig:profile_types}). In neither case is a significant mass dependence observed; notably, the separation by void type at mid-to-large void-centric distances is especially independent of mass\footnote{Similar results are obtained for stellar mass, but we choose not to show them here for simplicity. Note also that secondary bias is defined at fixed halo mass, thus the relevance of showing that potential dependence.}. The dependence of galaxy bias on void size and, particularly, void type thus occurs at the level of secondary bias, linking our results to the assembly bias effect (e.g., \citealt{ShethTormen2004,Gao2005,Paranjape2018,SatoPolito2019,Ramakrishnan2019, Ramakrishnan2020,Tucci2021,MonteroDorta2021, MonteroRodriguez2024,MonteroDorta2025}). These results are consistent with the simple model proposed by \cite{ShiSheth2018}, which describes the dependence of large-scale halo bias on both halo mass and environment\footnote{The authors explain why, at fixed environment, the dependence of halo bias on halo mass is expected to be weak.}. It would be interesting to extend the analysis of \cite{ShiSheth2018} to the case of the average void bias profiles studied here. In particular, it would be useful to determine whether this approach can predict the slope of the correlation shown in the right panel of Fig.~\ref{fig:average_bias}.

Finally, we have verified that our results are qualitatively robust against changes in void selection within the spherical overdensity method. As future work, we will evaluate how the galaxy bias profile inside voids may change when using alternative void finders, including modified spherical overdensity approaches such as the so-called popcorn voids \citep{Paz2023}, as well as fundamentally different techniques based on feature analysis of the matter density field \citep{Platen2007, Neyrinck2008, Sutter2015}, or methods utilizing the tidal or velocity fields \citep{Hahn2007, Lavaux2010, Elyiv2015}.

\section{Summary}
\label{sec:summary}

In this work, we leverage an object-by-object estimator of large-scale galaxy bias applied to the TNG300 hydrodynamical simulation to investigate, in unprecedented detail, the bias profile for central galaxies inside cosmic voids. In our study, voids are identified using a spherical overdensity technique. Our main conclusions can be summarized as follows:

\begin{itemize}
    \item The individual galaxy bias estimate isolates the contribution of each galaxy to the large-scale bias of a given population, which can be used to explore the bias conditioned on very specific regions of the cosmic web.
    
    \item The average large-scale bias of galaxies inside voids tends to increase outward, i.e., as a function of void-centric distance normalized by the void radius.
    
    \item The average large-scale bias of the entire (central) galaxy population inside voids increases with the overdensity parameter $\Delta_{\rm max}$, which represents the maximum overdensity of the surrounding environment, computed here within the range $[2R_{\rm void}, 3R_{\rm void}]$. Since this parameter is inversely correlated with void size, larger voids tend to host galaxy populations with lower bias.
    
    \item As a result of this dependence, the average large-scale galaxy bias inside S-type voids ($\Delta_{\rm max}\ge0$), which are embedded in large-scale overdense regions, is significantly higher at any void-centric distance than that of R-type voids ($\Delta_{\rm max}<0$), which, in contrast, are surrounded by underdense regions. The galaxy bias profile is also slightly steeper for S-type voids.
    
    \item R- and S-type voids contain halo and galaxy populations of similar mass. In analogy with the assembly bias effect, the difference in bias that we measure can be viewed as a secondary bias dependence. Similar conclusions can be drawn for the related size dependence.
    
    \item These conclusions remain robust under variations in the tracer set and void identification parameters.
\end{itemize}

Our results are expected to have multiple applications. 
First, the potential measurement of large-scale galaxy bias on an object-by-object basis in galaxy redshift surveys -- and its behaviour within voids -- can open new avenues for extracting cosmological information related to the growth of structures \citep{Hamaus2016, Correa2019,Correa2021,Correa2022,Song2024,contarini2024}. For example, a better understanding of the bias in and around cosmic voids could lead to improvements in the estimation of void velocity profiles within the Gaussian streaming model \citep{Fisher1995, DesjacquesSheth2010, Paz_clues_2013, Hamaus2015}, or enhance cosmological tests based on the abundance of voids identified using biased tracers \citep{Contarini2019, Paz2023}. 
Second, we will explore the universality of the bias profiles in voids with respect to modifications in the underlying cosmological model and the methods used to identify these underdense regions. 
Third, comparing our results with measurements in other LSS environments, such as clusters and filaments, will provide a more comprehensive understanding of how individual galaxy bias varies across different cosmic environments. With this in mind, we have recently generalized the object-by-object estimator (Eq. \ref{eq:bias}) for large-scale bias to estimate the large-scale tidal shear on an object-by-object basis.  We will report the results of this exercise elsewhere. Finally, our measurements can be incorporated into related analyses, such as the evolution of the Halo Occupation Distribution in voids, which exhibits distinct characteristics compared to other cosmic web environments (Alfaro et al., in prep), or the alignment of central galaxies in voids, which also appears to differ significantly from that in denser environments \citep{Rodriguez2025}.

\begin{acknowledgements} 

ADMD thanks Fondecyt for financial support through the Fondecyt Regular 2021
grant 1210612. ABA acknowledges the Spanish Ministry of Economy and
Competitiveness (MINECO) under the Severo Ochoa program SEV-2015-0548 grants.

ANR and FR thank the support by Agencia Nacional de Promoci\'on Cient\'ifica y
Tecnol\'ogica, the Consejo Nacional de Investigaciones Cient\'{\i}ficas y
T\'ecnicas (CONICET, Argentina) and the Secretar\'{\i}a de Ciencia y
Tecnolog\'{\i}a de la Universidad Nacional de C\'ordoba (SeCyT-UNC, Argentina).
ADMD and FR thank the ICTP for their hospitality and financial support through
the Senior Associates Programme 2022–2027 and Junior Associates Programme
2023–2028, respectively.

The IllustrisTNG simulations were undertaken with compute time awarded by the
Gauss Centre for Supercomputing (GCS) under GCS Large-Scale Projects GCS-ILLU
and GCS-DWAR on the GCS share of the supercomputer Hazel Hen at the High
Performance Computing Center Stuttgart (HLRS), as well as on the machines of the
Max Planck Computing and Data Facility (MPCDF) in Garching, Germany.

\end{acknowledgements}

%
%

\bibliographystyle{aa} 
\bibliography{references} 

\end{document}